\documentclass[aip,jcp,reprint,onecolumn,superscriptaddress]{revtex4}

\usepackage{amsmath}
\usepackage{mathtools}
\usepackage{amsbsy}
\usepackage{bm}
\usepackage{amssymb}
\usepackage{graphics}
\usepackage{graphicx}
\usepackage[colorlinks]{hyperref}
\usepackage{color}
\usepackage{subfig}
\usepackage{epstopdf}
\usepackage{esint}
\allowdisplaybreaks

\def\denM{{\mathcal{D}}}

\def\den{{\rho_{\mathcal{D}}}}
\def\bPsi{{\bm{\Psi}}}
\def\bg{{\mathbf{g}}}

\def\R{{\mathbb{R}}}

\def\bR{{\mathbf{R}}}

\def\bRJ{{\mathbf{R}_{J}}}
\def\bRJp{{\mathbf{R}_{J'}}}
\def\be{{\mathbf{x}}}

\def\bg{{\mathbf{g}}}

\def\Fab{{F\mkern-3mu_{\alpha \beta}}}
\def\sab{{\sigma \mkern-2mu_{\alpha \beta}}}
\def\grada{{\nabla \mkern-6mu_{x_{\alpha}}}}
\def\gradb{{\nabla \mkern-6mu_{x_{\beta}}}}
\def\gradya{{\nabla \mkern-6mu_{y_{\alpha}}}}
\def\gradyb{{\nabla \mkern-6mu_{y_{\beta}}}}
\def\gradx{{\bm{\nabla \mkern-6mu_{x}}}}

\def\R{{\mathbb{R}}}
\def\occ{\textsl{g}}

\def \gammaJl{\gamma_{_{Jl}}}
\def\tchins{{\tilde{\chi}_{_{Jlm}}}}

\def\pchins{{\chi_{_{J'lm}}}}

\def\pchis{{\chi_{_{J'lm}}}}

\begin{document}

\title{Real-space formulation of the stress tensor for $\mathcal{O}(N)$ density functional theory: application to high temperature calculations}
\author{Abhiraj Sharma}
\affiliation{College of Engineering, Georgia Institute of Technology, GA 30332, USA}
\author{Sebastien Hamel}
\affiliation{Physics Division, Lawrence Livermore National Laboratory, Livermore, CA 94550, USA}
\author{Mandy Bethkenhagen}
\affiliation{Physics Division, Lawrence Livermore National Laboratory, Livermore, CA 94550, USA}
\affiliation{CNRS, École Normale Supérieure de Lyon, Laboratoire de Géologie de Lyon LGLTPE UMR5276, Centre Blaise Pascal, 46 allée d’Italie Lyon 69364, France}
\author{John E. Pask}
\affiliation{Physics Division, Lawrence Livermore National Laboratory, Livermore, CA 94550, USA}
\author{Phanish Suryanarayana}
\email[Email: ]{phanish.suryanarayana@ce.gatech.edu}
\affiliation{College of Engineering, Georgia Institute of Technology, GA 30332, USA}
\date{\today}

\begin{abstract}
We present an accurate and efficient real-space formulation of the Hellmann-Feynman stress tensor for $\mathcal{O}(N)$ Kohn-Sham density functional theory (DFT). While applicable at any temperature, the formulation is most efficient at high temperature where the Fermi-Dirac distribution becomes smoother and density matrix becomes correspondingly more localized. We first rewrite the orbital-dependent stress tensor for real-space DFT in terms of the density matrix, thereby making it amenable to $\mathcal{O}(N)$ methods. We then describe its evaluation within the $\mathcal{O}(N)$ infinite-cell Clenshaw-Curtis Spectral Quadrature (SQ) method, a technique that is applicable to metallic as well as insulating systems, is highly parallelizable, becomes increasingly efficient with increasing temperature, and provides results corresponding to the infinite crystal without the need of Brillouin zone integration. We demonstrate systematic convergence of the resulting formulation with respect to SQ parameters to exact diagonalization results, and show convergence with respect to mesh size to established planewave results. We employ the new formulation to compute the viscosity of hydrogen at a million kelvin from Kohn-Sham quantum molecular dynamics, where we find agreement with previous more approximate orbital-free density functional methods.  
\end{abstract}
\maketitle

\section{Introduction}
Kohn-Sham density functional theory (DFT) \cite{Hohenberg,Kohn1965} is among the most widely used first principles methods for understanding and predicting materials properties. The tremendous popularity of DFT can be attributed to its favorable accuracy-to-cost ratio and simplicity relative to other such ab-initio theories. In DFT simulations for condensed matter systems, a fundamental quantity of interest, in addition to the ground-state energy and atomic forces, is the second order Hellmann-Feynman stress tensor. The components  of this tensor represent derivatives of the energy density with respect to the six independent homogeneous strains that can be applied to the system, evaluated at the electronic ground state. The stress tensor has many applications, including the determination of equilibrium lattice parameters, equation of state (EOS), and the calculation of shear viscosity from quantum molecular dynamics (QMD) simulations \cite{de1998viscosity,jakse2013liquid,PhysRevB.99.165103}.

The Hellmann-Feynman stress tensor in DFT, which has its origins in the work of Slater \cite{Slater} and Janak \cite{Janak}, has been developed in the context of pseudopotential planewave calculations \cite{Yin,Nielsen1,Nielsen2,StressGGA1994}, the linearized augmented plane wave (LAPW) method \cite{Thonhauser2002}, the projector augmented-wave (PAW) method \cite{PAWABINIT2008}, atom-centered orbital bases \cite{SIESTA,Knuth2015}, the finite-element method \cite{Motamarriconfigurationalforce}, and the  pseudopotential real-space finite-difference method \cite{sharma2018calculation}. However, a common feature of all these formulations is their expression in terms of Kohn-Sham orbitals, the calculation of which is associated with computational cost and memory requirements that scale as $\mathcal{O}(N^3)$ and $\mathcal{O}(N^2)$ with respect to the number of atoms, respectively. This critical cubic scaling arises due to the orthonormality constraint on the Kohn-Sham orbitals, which also limits scalability in the context of high performance parallel computing, severely limiting the length and time scales that can be reached.

In order to overcome the critical $\mathcal{O}(N^3)$ bottleneck, a number of $\mathcal{O}(N)$ approaches have been developed (e.g., \cite{Goedecker,Bowler2012,aarons2016perspective} and references therein) which circumvent the calculation of the Kohn-Sham orbitals by proceeding instead through the density matrix to determine quantities of interest, achieving linear scaling by exploiting the exponential decay of the density matrix for insulating systems and metallic systems at finite temperature \cite{goedecker1998decay,ismail1999locality,zhang2001properties,taraskin2002spatial,benzi2013decay}. These efforts have culminated in a number of mature codes \cite{SIESTA,Conquestref,ONETEP,FEMTECKref,MGMolref,BigDFTref,OpenMXweb,FreeONref,aarons2018electronic,mohr2018linear}, however important challenges remain. These include limitations of underlying basis sets, large prefactors, the need for additional computational parameters, subtleties in determining sufficient numbers and/or centers of localized orbitals, the calculation of accurate atomic forces, and large-scale parallelization \cite{Bowler2012,RuiHinSky12}. In particular, to the best of our knowledge, the stress tensor has not been formulated and implemented in the context of $\mathcal{O}(N)$ DFT, especially for calculations at high temperature. 

Kohn-Sham calculations at high temperature occur in a range of applications areas, including the study of warm dense matter and dense plasmas, as occur in laser experiments and the interiors of giant planets and stars \cite{gradesred2014,grabasben2011,renaudin2003aluminum,dharma2006static,ernstorfer2009formation,white2013orbital}. However, these calculations present significant challenges due to the substantially larger number and lesser locality of orbitals that must be computed. Consequently, $\mathcal{O}(N^3)$  as well as local-orbital based $\mathcal{O}(N)$ methods have very large prefactors, which makes QMD calculations for even small systems intractable. The recently developed Spectral Quadrature (SQ) method \cite{suryanarayana2013spectral,pratapa2016spectral,suryanarayana2017sqdft} addresses scaling with number of atoms as well as temperature, while retaining systematic convergence to standard diagonalization results for metals and insulators alike.  In particular, due to the increased locality of electronic interactions and enhanced smoothness of the Fermi-Dirac function, the cost of the $\mathcal{O}(N)$ SQ approach decreases rapidly as temperature is increased. Furthermore, it is particularly well suited to massive parallelization since a majority of the communication is localized to nearby processors, the pattern of which remains fixed throughout the calculation. 

In this work, we present an accurate and efficient real-space formulation of the stress tensor for  $\mathcal{O}(N)$ Kohn-Sham calculations. While applicable at any temperature, the formulation is most efficient at high temperature where the Fermi-Dirac distribution becomes smoother and density matrix becomes correspondingly more localized. Starting with the recently derived expression for the stress tensor in real-space DFT  \cite{sharma2018calculation}, we develop a formulation in terms of the density matrix, and then describe its evaluation within the infinite-cell Clenshaw-Curtis SQ method. The framework is applicable to metallic as well as insulating systems, is highly parallelizable, becomes more efficient as the temperature is increased, and stresses corresponding to the infinite crystal can be computed without Brillouin zone integration. We demonstrate the systematic convergence of the stress tensor with respect to quadrature order as well as truncation radius---the two key parameters of the SQ method in addition to mesh size---to exact diagonalization results. In addition, we show convergence with respect to mesh size to established planewave results. Finally, we employ the new formulation to compute the viscosity of hydrogen at a million kelvin from Kohn-Sham QMD.

The remainder of this manuscript is organized as follows. In Section~\ref{Sec:ON3Stress}, we summarize the  orbital-dependent stress tensor formulation for real-space DFT. Next, we develop a density matrix formulation for the stress tensor in Section~\ref{Sec:ONStress} and describe its evaluation using the real-space SQ method in Section~\ref{Sec:StressSQ}. Finally, we verify the proposed framework in Section~\ref{Sec:Results}, and provide concluding remarks in Section~\ref{Sec:Conclusions}. 


\section{Stress tensor in $\mathcal{O}(N^3)$ real-space DFT} \label{Sec:ON3Stress}
\begin{figure}[h!]
\includegraphics[keepaspectratio=true,width=0.35\textwidth]{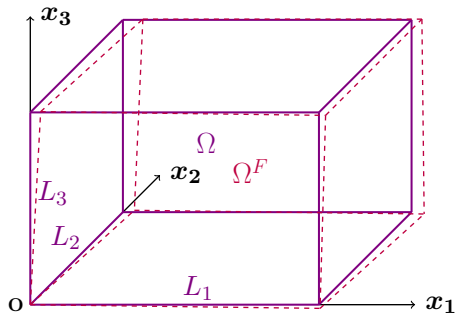}
\caption{\label{Fig:celldeformation} Orthorhombic unit cell $\Omega$ (solid lines) with sides of length $L_1$, $L_2$, and $L_3$ along the $\bm{x_1}$, $\bm{x_2}$, and $\bm{x_3}$ directions, respectively. The unit cell subsequent to the application of the infinitesimal homogeneous deformation is denoted by $\Omega^F$ (dashed lines).}
\end{figure}

Consider an orthorhombic unit cell $\Omega$ (Fig.~\ref{Fig:celldeformation}), with nuclei positioned at $\bR = \{\bR_1,\bR_2, \ldots ,\bR_N\}$ and a total of $N_e$ valence electrons. The lattice vectors corresponding to $\Omega$ are $L_1 \bm{\hat{e}_1}$, $L_2 \bm{\hat{e}_2}$, and $L_3 \bm{\hat{e}_3}$, where $\bm{\hat{e}_1}$, $\bm{\hat{e}_2}$, and $\bm{\hat{e}_3}$ are the lattice/Cartesian unit vectors. For an infinitesimal homogeneous deformation that maps $\Omega$ to $\Omega^{F}$, as illustrated in Fig.~\ref{Fig:celldeformation}, the Hellmann-Feynman stress tensor can be defined as \cite{sharma2018calculation}
\begin{equation} \label{Eqn:stressdefinition}
\sab = \frac{1}{|\Omega|} \frac{\partial \mathcal{L}^{F}(\bPsi,\bg,\phi,\bR \mkern-2mu^{F})}{\partial \Fab}\Bigg|_{\mathcal{G}} \,, \quad \alpha, \beta \in \{1, 2, 3\} \,,  
\end{equation}
where $|\Omega|$ denotes the measure of the unit cell, the superscript $(.)^F$ represents quantities after perturbation of the unit cell with a deformation gradient having components $\Fab$, $\mathcal{L}$ denotes the Lagrangian, $\bPsi = \{\psi_1,\psi_2, \ldots ,\psi_{N_s}\}$ is the collection of Kohn-Sham orbitals with occupations $\bg = \{\occ_1,\occ_2, \ldots \occ_{N_s}\}$, $\phi$ is the electrostatic potential \cite{Phanish2010,suryanarayana2011mesh}, and $\mathcal{G}$ signifies the electronic ground state for the undeformed system. 

Neglecting spin and Brillouin zone integration, the real-space stress tensor for the choice of a semilocal exchange-correlation functional and a local formulation of the electrostatics \cite{pask2005real,Suryanarayana2014524,ghosh2016higher} can be written as \cite{sharma2018calculation}:
\begin{equation}
\sab = \frac{1}{|\Omega|} \bigg[\sigma \mkern-2mu ^{I}_{\alpha \beta} + \sigma \mkern-2mu ^{II}_{\alpha \beta} + \sigma \mkern-2mu ^{III}_{\alpha\beta} + \sigma \mkern-2mu ^{IV}_{\alpha \beta}  \bigg] \,, \label{Eqn:stressexpansion}
\end{equation}
where $\sigma \mkern-2mu ^{I}_{\alpha \beta}$, $\sigma \mkern-2mu ^{II}_{\alpha \beta}$, $\sigma \mkern-2mu ^{III}_{\alpha \beta}$, and $\sigma \mkern-2mu ^{IV}_{\alpha \beta}$ are the contributions arising from the electronic kinetic energy $T_s$, exchange-correlation energy $E_{xc}$, nonlocal pseudopotential energy $E_{nl}$, and the electrostatic energy $E_{el}$, respectively:  
\begin{eqnarray}
\sigma \mkern-2mu ^{I}_{\alpha \beta} & = &   -2 \sum_{n=1}^{N_s} \occ_n \int_\Omega \grada \psi_n(\be) \gradb \psi_n(\be) \, \mathrm{d \be}  \,, \label{Eq:sigmaI} \\
\sigma \mkern-2mu ^{II}_{\alpha \beta} & = & \delta_{\alpha \beta} \bigg( E_{xc}(\rho,\bm{\nabla} \rho) - \int_\Omega V_{xc}\big(\rho(\be),\gradx \rho(\be)\big) \rho(\be) \, \mathrm{d \be} \bigg) - \int_\Omega \rho(\be) \frac{\partial \varepsilon_{xc}\big(\rho(\be),\gradx \rho(\be) \big)}{\partial \big(\gradb \rho(\be)\big)} \grada \rho(\be) \, \mathrm{d \be} \,, \label{Eq:sigmaII} \\
\sigma \mkern-2mu ^{III}_{\alpha \beta} & = & -\delta_{\alpha \beta} E_{nl}(\bPsi,\bg,\bR) \nonumber \\
& & - 4 \sum_{n=1}^{N_s}  \occ_n \sum_J \sum_{lm} \gammaJl  \Bigg(\sum_{J'} \int_\Omega \pchis(\be,\bRJp){\big(\be - \bRJp \big)\mkern-4mu}_\beta \grada \psi_n(\be)  \, \mathrm{d \be} \Bigg) \Bigg(\int_\Omega \tchins(\be,\bRJ) \psi_n(\be) \, \mathrm{d \be}\Bigg)  \,, \label{Eq:sigmaIII} \\
\sigma \mkern-2mu ^{IV}_{\alpha \beta} & = & \frac{1}{4 \pi} \int_\Omega \grada \phi(\be,\bR) \gradb \phi(\be,\bR) \, \mathrm{d \be} + \sum_I \int_\Omega \grada b_I(\be,\bR_I) {\big(\be - \bR_I \big)\mkern-4mu}_\beta \Big(\phi(\be,\bR) - \frac{1}{2} \, V_I(\be,\bR_I) \Big) \, \mathrm{d \be}\nonumber \\
&-& \frac{1}{2} \sum_I \int_\Omega \grada V_I(\be,\bR_I) {\big(\be - \bR_I \big)\mkern-4mu}_\beta b_I(\be,\bR_I) \, \mathrm{d \be} + \frac{1}{2} \delta_{\alpha \beta} \int_\Omega \big(b(\be,\bR) - \rho(\be) \big) \phi(\be,\bR) \, \mathrm{d \be} -\delta_{\alpha \beta} E_{self}(\bR) \nonumber \\
&+ &  \sigma \mkern-2mu ^{E_c}_{\alpha\beta}  \,. \label{Eq:sigmaIV}
\end{eqnarray}
The electron density $\rho$ can itself be expanded as
\begin{equation} \label{Eqn:electrondensity}
\rho(\be) = 2 \sum_{n=1}^{N_s}\occ_n \psi_n^2(\be)  \,,
\end{equation}
the electrostatic potential $\phi$ is the solution of the Poisson equation
\begin{equation}
-\frac{1}{4 \pi} \nabla^2 \phi (\be,\bR) = \rho(\be,\bR) + b(\be,\bR) \,,
\end{equation}
and the energy terms $E_{xc}$ and $E_{nl}$ take the form
\begin{eqnarray}
E_{xc} (\rho,\bm{\nabla} \rho) & = & \int_{\Omega} \varepsilon_{xc} \big(\rho(\be),\gradx \rho(\be)\big) \rho(\be) \, \mathrm{d \be} \,, \\
E_{nl}(\bPsi,\bg,\bR) & = & 2 \sum_{n=1}^{N_s} \occ_n \sum_J \sum_{lm} \gammaJl \bigg ( \int_{\Omega} \tchins(\be,\bRJ) \psi_n(\be) \, \mathrm{d\be} \bigg)^2  \,.
\end{eqnarray}
In the above equations, $\grada$ is the $\alpha^{th}$ component of the gradient vector $\gradx$; $\delta_{\alpha \beta}$ is the Kronecker delta function; $\lambda_n$ are the eigenvalues corresponding to $\psi_n$; $\varepsilon_{xc}$ is the sum of the exchange and correlation energy per particle of a uniform electron gas, $V_{xc}$ is the exchange-correlation potential; $\tchins= \sum_{J'} \pchins $ are the periodically mapped nonlocal projectors, with $\pchins$ representing the projectors associated with the $J'^{th}$ atom and the index $lm$ running over all azimuthal and magnetic quantum numbers; $b= \sum_{I} b_I$ is the total pseudocharge density of the nuclei, with $b_I$ being the pseudocharge density of the $I^{th}$ nucleus that generates the potential $V_I$; $E_{self}=\frac{1}{2}\sum_I \int_{\Omega} b_I(\be,\bR_I) V_I(\be,\bR_I) \, \mathrm{d\be}$ is the self energy associated with the pseudocharge densities; $\sigma \mkern-2mu ^{E_c}_{\alpha\beta}$ is the stress tensor contribution arising from the energy correction due to the overlapping pseudocharges \cite{sharma2018calculation}; the summation index $J$ runs over all atoms in $\Omega$; the summation index $J'$ runs over the $J^{th}$ atom and its periodic images; and the summation index $I$ runs over all atoms in $\R^3$. It is worth noting that all quantities in the above equations correspond to the electronic ground state, i.e., after the solution of the Kohn-Sham nonlinear eigenproblem. 


\section{Stress tensor in $\mathcal{O}(N)$ real-space DFT} \label{Sec:ONStress}
The Hellmann-Feynman stress tensor presented in the previous section has been formulated in terms of the Kohn-Sham orbitals $\psi_n$. Though the evaluation of the stress tensor by itself scales as $\mathcal{O}(N)$, calculation of the orbitals  scales as $\mathcal{O}(N^3)$ with respect to system size \cite{Martin2004}. Since $\mathcal{O}(N)$ methods \cite{Goedecker,Bowler2012,aarons2016perspective} bypass the calculation of the orbitals,  computing the truncated density matrix (directly or indirectly) instead, the stress tensor needs to be suitably reformulated in terms of this quantity. To do so, we note that the density matrix is of the form
\begin{equation}
\mathcal{D}(\be,\mathbf{y}) = \sum_{n=1}^{N_s} g_n \psi_n(\be) \psi_n(\mathbf{y}) \,,
\end{equation}
whose diagonal entries are related to the electron density by 
\begin{equation} \label{Eqn:DMelectrondensity}
\den(\be) = 2 \mathcal{D}(\be,\be) \,. \\
\end{equation} 
Indeed, the density matrix has exponential decay for insulators as well as metals at finite electronic temperature \cite{goedecker1998decay,benzi2013decay}, allowing for its truncation during computation. 

Among the various stress tensor contributions in Eqns.~\ref{Eq:sigmaI}--\ref{Eq:sigmaIV},  only $\sigma \mkern-2mu ^{I}_{\alpha \beta}$ and $\sigma \mkern-2mu ^{III}_{\alpha \beta}$ have an explicit dependence on the orbitals. Therefore, we now rewrite them in terms of the density matrix:
\begin{eqnarray}
\sigma \mkern-2mu ^{I}_{\alpha \beta} & = &   -2 \sum_{n=1}^{N_s} \occ_n \int_\Omega \grada \psi_n(\be) \gradb \psi_n(\be) \, \mathrm{d \be} \nonumber \\
& = & 2 \sum_{n=1}^{N_s} \occ_n \int_\Omega \psi_n(\be) \grada \gradb \psi_n(\be) \, \mathrm{d \be} \nonumber \\ 
& = & 2 \int_\Omega  \bigg( \gradya \gradyb \sum_{n=1}^{N_s} \occ_n \psi_n(\mathbf{y}) \psi_n(\be) \bigg)\bigg|_{\mathbf{y} = \be} \, \mathrm{d \be} \nonumber \\
& = & 2 \int_\Omega  \bigg( \gradya \gradyb \mathcal{D}(\mathbf{y},\be)\bigg)\bigg|_{\mathbf{y} = \be} \, \mathrm{d \be} \,, \label{Eq:stressIRe}
\end{eqnarray}
and
\begin{eqnarray}
\sigma \mkern-2mu ^{III}_{\alpha \beta} & = & -\delta_{\alpha \beta} \, 2 \sum_{n=1}^{N_s} \occ_n \sum_J \sum_{lm} \gammaJl \bigg( \int_\Omega \tchins(\be,\bRJ) \psi_n(\be)  \, \mathrm{d \be} \bigg) \bigg(\int_\Omega \tchins(\mathbf{y},\bRJ) \psi_n(\mathbf{y}) \, \mathrm{d \mathbf{y}} \bigg) \nonumber \\
& - & 4 \sum_{n=1}^{N_s} \occ_n \sum_J \sum_{lm} \gammaJl \Bigg(\sum_{J'} \int_\Omega \pchis(\be,\bRJp) {\big(\be - \bRJp \big)\mkern-4mu}_\beta \grada \psi_n(\be)  \, \mathrm{d \be} \Bigg)\Bigg(\int_\Omega \tchins(\mathbf{y},\bRJ) \psi_n(\mathbf{y}) \, \mathrm{d \mathbf{y}}\Bigg) \nonumber \\
& = & -\delta_{\alpha \beta} \, 2 \sum_J \sum_{lm} \gammaJl \bigg( \int_\Omega \int_\Omega \tchins(\be,\bRJ) \sum_{n=1}^{N_s} \occ_n \psi_n(\be) \psi_n(\mathbf{y}) \tchins(\mathbf{y},\bRJ) \, \mathrm{d \be} \, \mathrm{d \mathbf{y}} \bigg) \nonumber \\
& - & 4 \sum_J \sum_{lm} \gammaJl \sum_{J'} \int_\Omega \int_\Omega \pchis(\be,\bRJp) {\big(\be - \bRJp \big)\mkern-4mu}_\beta \grada \sum_{n=1}^{N_s} \occ_n \psi_n(\be) \psi_n(\mathbf{y}) \tchins(\mathbf{y},\bRJ) \, \mathrm{d \be}  \, \mathrm{d \mathbf{y}}  \nonumber \\
& = & -\delta_{\alpha \beta} \, 2 \sum_J \sum_{lm} \gammaJl \bigg( \int_\Omega \int_\Omega \tchins(\be,\bRJ) \mathcal{D}(\be, \mathbf{y}) \tchins(\mathbf{y},\bRJ) \, \mathrm{d \be} \, \mathrm{d \mathbf{y}} \bigg) \nonumber \\
& - & 4 \sum_J \sum_{lm} \gammaJl  \sum_{J'} \int_\Omega \int_\Omega \pchis(\be,\bRJp) {\big(\be - \bRJp \big)\mkern-4mu}_\beta \grada \mathcal{D}(\be, \mathbf{y}) \tchins(\mathbf{y},\bRJ) \, \mathrm{d \be}  \, \mathrm{d \mathbf{y}} \,. \label{Eq:stressIIIRe}
\end{eqnarray}
The second equality for $\sigma \mkern-2mu ^{I}_{\alpha \beta}$ is obtained using integration by parts in conjunction with the divergence theorem, whereas the third equality is obtained by a rearrangement of terms. The second as well as the third equalities for $\sigma \mkern-2mu ^{III}_{\alpha \beta}$ are obtained by rearrangement of terms. 

Using the above relations and Eqns.~\ref{Eqn:stressexpansion}, \ref{Eq:sigmaII}, and \ref{Eq:sigmaIV}, 
we arrive at the following reformulation for the Hellmann-Feynman stress tensor in terms of the density matrix:
\begin{eqnarray}
\sigma \mkern-2mu_{\alpha \beta} &=& \frac{1}{|\Omega |} \Bigg[ 2  \int_\Omega  \bigg( \gradya \gradyb \mathcal{D}(\mathbf{y},\be)\bigg)\bigg|_{\mathbf{y} = \be} \, \mathrm{d \be} + \delta_{\alpha \beta} \bigg( E_{xc}(\rho_\mathcal{D}, \bm{\nabla} \rho_\mathcal{D}) - \int_\Omega V_{xc}\big(\rho_\mathcal{D}(\be), \gradx \rho_\mathcal{D}(\be)\big) \rho_\mathcal{D}(\be) \, \mathrm{d \be} \bigg) \, \nonumber \\
& - &\int_\Omega \rho_\mathcal{D}(\be) \frac{\partial \varepsilon_{xc}\big(\rho_\mathcal{D}(\be),\gradx \rho_\mathcal{D}(\be) \big)}{\partial \big(\gradb \rho_\mathcal{D}(\be)\big)} \grada \rho_\mathcal{D}(\be) \, \mathrm{d \be} - \delta_{\alpha \beta} \, E_{nl}(\denM,\bR)  \nonumber \\
& - & \,4 \, \sum_J \sum_{lm} \gammaJl \, \sum_{J'} \int_\Omega \int_\Omega \pchis(\be,\bRJp) {\big(\be - \bRJp \big)\mkern-4mu}_\beta \grada D(\be,\mathbf{y}) \tchins(\mathbf{y},\bRJ) \, \mathrm{d \be} \, \mathrm{d \mathbf{y}}   \nonumber + \frac{1}{4 \pi} \int_{\Omega} \grada \phi(\be,\bR) \gradb \phi(\be,\bR) \, \mathrm{d \be} \nonumber \\
&+& \sum_I \int_\Omega \grada b_I(\be,\bR_I) {\big(\be - \bR_I \big)\mkern-4mu}_\beta \Big(\phi(\be,\bR) - \frac{1}{2} \, V_I(\be,\bR_I) \Big) \, \mathrm{d \be} - \frac{1}{2} \sum_I \int_\Omega \grada V_I(\be,\bR_I) {\big(\be - \bR_I \big)\mkern-4mu}_\beta b_I(\be,\bR_I) \, \mathrm{d \be} \nonumber \\
& + &   \frac{1}{2} \,\delta_{\alpha \beta} \int_\Omega \big(b(\be,\bR) - \rho_{\mathcal{D}}(\be) \big) \phi(\be,\bR) \, \mathrm{d \be} - \delta_{\alpha \beta} \, E_{self}(\bR) + \sigma \mkern-2mu ^{E_c}_{\alpha\beta} \Bigg] \,. \label{Eqn:ON:Stress}
\end{eqnarray}
The evaluation of the stress tensor by itself scales as $\mathcal{O}(N)$ with respect to system size by virtue of the truncated nature of the density matrix. The overall scaling also becomes $\mathcal{O}(N)$  when a linear-scaling method is used to calculate the truncated density matrix.  Note that, proceeding along the lines of Ref.~\cite{sharma2018calculation}, it is straightforward to generalize the above formulation to include  non-orthogonal systems and Brillouin zone integration. However, we focus on the orthogonal case here for simplicity since non-orthogonal cells and Brillouin zone integration are typically not required in $\mathcal{O}(N)$ calculations, where target systems tend to be large. In addition, the stresses corresponding to the infinite crystal can be computed in the Spectral Quadrature (SQ) formalism \cite{suryanarayana2013spectral,pratapa2016spectral,suryanarayana2017sqdft} without Brillouin zone integration.


\section{Stress tensor in $\mathcal{O}(N)$ real-space Spectral Quadrature method} \label{Sec:StressSQ}

We now describe the evaluation of the stress tensor within the framework of the $\mathcal{O}(N)$ real-space Spectral Quadrature (SQ) method \cite{suryanarayana2013spectral,pratapa2016spectral,suryanarayana2017sqdft}. In this approach, all quantities of interest (in discrete form) are expressed as bilinear forms or sums of bilinear forms, which are then approximated by spatially localized quadrature rules. The method does not assume the existence of a band gap and so is applicable to metallic and insulating systems alike. In addition, the technique is  particularly well suited to scalable high performance computing  and becomes more efficient as the temperature is increased, by virtue of the  electronic interactions becoming more localized and the representation of the Fermi-Dirac function becoming more compact.

It is clear from Eqn.~\ref{Eqn:ON:Stress}  that the off-diagonal components of the density matrix are required for calculation of the stress tensor. In such a situation, Clenshaw-Curtis SQ is significantly more efficient than the Gauss SQ variant, motivating its selection here \cite{pratapa2016spectral}. In particular, we choose the  \emph{infinite-cell} version of Clenshaw-Curtis SQ,  wherein results corresponding to the infinite crystal are obtained without recourse to Brillouin zone integration or large supercells. Specifically, rather than employ Bloch boundary conditions for the orbitals on the unit cell $\Omega$, zero-Dirichlet (or equivalently periodic) boundary conditions are prescribed at infinity, and the relevant components  of the density matrix for spatial points within $\Omega$ are calculated using the nearsightedness principle \cite{prodan2005nearsightedness}. Indeed, the infinite-cell approach reduces to  the standard $\Gamma$-point calculation when the size of the truncation region is smaller than the domain size, a situation common in large-scale $\mathcal{O}(N)$ DFT simulations, particularly those at high temperature. 

Proceeding as in previous work \cite{pratapa2016spectral,suryanarayana2017sqdft}, we discretize the domain $\Omega$ with a uniform grid containing $N_d$ finite-difference nodes, the collection of which is denoted by $K_\Omega$. Then we partition $\Omega$ into $N_p$ non-overlapping regions of equal size such that $\Omega = \bigcup^{N_p}_{p=1} \Omega_p$ and $K_\Omega = \bigcup_{p=1}^{N_p} K_{\Omega_p}$, where $N_p$ is the total number of processors and $K_{\Omega_p}$ denotes the collection of grid points associated with $\Omega_p$, the domain local to the $p^{th}$ processor. We define the nodal Hamiltonian $\mathbf{H}_q \in \R^{N_c \times N_c}$ for node $q \in K_\Omega$ as the restriction of the Hamiltonian to its \emph{region of influence}---the cuboid of side $2 R_{cut}$ centered at that point containing a total of $N_c$ finite-difference nodes, with $R_{cut}$ denoting the truncation radius of the density matrix beyond which the electronic interactions are ignored. Similarly, we define $\mathbf{w}_q \in \R^{N_c \times 1}$, $\mathbf{\nabla}_{h,q} \equiv \left(\mathbf{\nabla}_{1 h,q} \in \R^{N_c \times N_c}, \, \mathbf{\nabla}_{2 h,q} \in \R^{N_c \times N_c}, \, \mathbf{\nabla}_{3 h,q} \in \R^{N_c \times N_c} \right)$, $\mathbf{V}^{I}_{nl,q} \in \R^{N_c \times N_c}$, and $\mathbf{X}_{h,q} \equiv \left( \mathbf{X}_{1 h,q} \in \R^{N_c \times N_c}, \, \mathbf{X}_{2 h,q} \in \R^{N_c \times N_c}, \, \mathbf{X}_{3 h,q} \in \R^{N_c \times N_c} \right)$ for node $q \in K_\Omega$ as the restriction to its region of influence of the standard basis vector, gradient matrices, nonlocal pseudopotential matrix of the $I^{th}$ atom, and the spatial location matrices of the grid points, respectively.

In the aforedescribed framework, the stress tensor contributions that explicitly depend on the density matrix, i.e., $\sigma \mkern-2mu ^{I}_{\alpha \beta}$ and $\sigma \mkern-2mu ^{III}_{\alpha \beta}$, can be written as follows:
\begin{eqnarray}
\sigma \mkern-2mu ^{I}_{\alpha \beta} &\approx & 2 \sum_{p=1}^{N_p} \sum_{q \in K_{\Omega_p}} \mathbf{w}_q^T \Big( \nabla \mkern-4mu_{\alpha h, q} \nabla \mkern-4mu_{\beta h, q} \sideset{}{'} \sum_{j=0}^{n_{pl}} c_q^j T_j(\hat{\mathbf{H}}_q) \Big) \mathbf{w}_q  \nonumber \\
&=& 2 \sum_{p=1}^{N_p} \sum_{q \in K_{\Omega_p}} \mathbf{w}_q^T \nabla \mkern-4mu_{\alpha h, q} \nabla \mkern-4mu_{\beta h, q} \bigg(\sideset{}{'} \sum_{j=0}^{n_{pl}} c_q^j \mathbf{t}_q^j \bigg) \,, \\
\sigma \mkern-2mu ^{III}_{\alpha \beta} &\approx & -2 \, \delta_{\alpha \beta} \sum_{p=1}^{N_p} \sum_{I \in D^c_{p'}} \sum_{q \in K_{\Omega_p}} \mathbf{w}_q^T \Big( V_{nl,q}^I \sideset{}{'} \sum_{j=0}^{n_{pl}} c_q^j T_j(\hat{\mathbf{H}}_q) \Big) \mathbf{w}_q  \nonumber \\
& - & 4 \sum_{p=1}^{N_p} \sum_{I \in D^c_{p'}} \sum_{q \in K_{\Omega_p}} \mathbf{w}_q^T \Big( V_{nl,q}^I \big( \mathbf{X}_{\beta h,q} - \bR_{I,\beta} \, \mathbf{I} \big) \nabla \mkern-4mu_{\alpha h, q} \sideset{}{'} \sum_{j=0}^{n_{pl}} c_q^j T_j(\hat{\mathbf{H}}_q) \Big) \mathbf{w}_q \nonumber \\
&=& -2 \sum_{p=1}^{N_p} \sum_{I \in D^c_{p'}} \sum_{q \in K_{\Omega_p}} \mathbf{w}_q^T V_{nl,q}^I \Bigg(\delta_{\alpha \beta} \bigg( \sideset{}{'} \sum_{j=0}^{n_{pl}} c_q^j \mathbf{t}_q^j \bigg) + 2 \, \big(\mathbf{X}_{\beta h,q} - \bR_{I,\beta} \, \mathbf{I} \big)  \nabla \mkern-4mu_{\alpha h, q} \bigg(\sideset{}{'} \sum_{j=0}^{n_{pl}} c_q^j \mathbf{t}_q^j \bigg) \Bigg) \,, 
\end{eqnarray} 
where $n_{pl}$ is the order of the Clenshaw-Curtis quadrature, $T_j$ denotes the Chebyshev polynomial of degree $j$; $\hat{\mathbf{H}}_q = ( \mathbf{H}_q - \chi_q \mathbf{I} ) / \zeta_q$ is the scaled and shifted nodal Hamiltonian having spectrum in $[ -1,1 ]$, with $\mathbf{I} \in \R^{N_c \times N_c}$ signifying the identity matrix, $\chi_q = (\lambda_q^{max} + \lambda_q^{min})/2$, and $\zeta_q = (\lambda_q^{max} - \lambda_q^{min})/2$, where $\lambda_q^{max}$ and $\lambda_q^{min}$ denote the maximum and minimum eigenvalues of $\mathbf{H}_q$, respectively; $D^c_{p'}$ is the set of all atoms in $\R^3$ whose nonlocal projectors have overlap with the extended processor domain, i.e., processor domain extended by $R_{cut}$ on each side; the summation with a prime indicates that the first term is halved;  $c_q^j$ is the Chebyshev expansion coefficient of the Fermi-Dirac function
\begin{eqnarray}
c_q^j = \frac{2}{\pi} \int_{-1}^{1} \frac{g(r,\hat{\mu}_q,\hat{\sigma}_q) T_j (r)}{\sqrt{1-r^2}} {\rm dr} \,, 
\end{eqnarray}
where $\hat{\mu}_q = (\mu - \chi_q)/ \zeta_q$ is the scaled and shifted Fermi energy and $\hat{\sigma} = \sigma/ \zeta_q$ is the scaled smearing; and $\mathbf{t}_q^j \in \R^{N_c \times 1}$ 
is the $q^{th}$ column of $T_j(\hat{\mathbf{H}}_q)$, determined using the three term recurrence relation for Chebyshev polynomials:
\begin{eqnarray}
\mathbf{t}_q^{i+1} &=& 2 \hat{\mathbf{H}}_q \mathbf{t}_q^{i} - \mathbf{t}_q^{i-1}, \quad i=1,2,\ldots,n_{pl} \nonumber \\
\mathbf{t}_q^1 &=& \hat{\mathbf{H}}_q \mathbf{w}_q \,, \, \mathbf{t}_q^0 = \mathbf{w}_q \,. 
\end{eqnarray}

Thereafter, the  expression for the stress tensor in Eqn.~\ref{Eqn:ON:Stress} take the form: 
\begin{eqnarray}
\sigma \mkern-2mu_{\alpha \beta} &=& \frac{1}{|\Omega |} \Bigg[|\mathrm{d\Omega|} \sum_{p=1}^{N_p} \sum_{q \in K_{\Omega_p}} \Bigg( \frac{2}{|\mathrm{d\Omega}|} \mathbf{w}_q^T \nabla \mkern-4mu_{\alpha h, q} \nabla \mkern-4mu_{\beta h, q} \bigg(\sideset{}{'} \sum_{j=0}^{n_{pl}} c_q^j \mathbf{t}_q^j \bigg) + \delta_{\alpha \beta} \Big(\epsilon_{xc}(\rho_q) \rho_q - V_{xc}(\rho_q) \rho_q \Big) - \rho_q \frac{\partial \varepsilon_{xc}\big(\rho,\bm \nabla_{h} \rho \big)}{\partial \big(\nabla_{\beta h} \rho \big)} \Bigg|_q \nabla_{\alpha h} \rho \Big|_q \Bigg) \nonumber \\ 
& - & 2 \sum_{p=1}^{N_p} \sum_{I \in D^c_{p'}} \sum_{q \in K_{\Omega_p}} \mathbf{w}_q^T V_{nl,q}^I \Bigg(\delta_{\alpha \beta} \bigg(\sideset{}{'} \sum_{j=0}^{n_{pl}} c_q^j \mathbf{t}_q^j \bigg) + 2 \, \big(\mathbf{X}_{\beta h,q} - \bR_{I,\beta} \, \mathbf{I} \big)  \nabla \mkern-4mu_{\alpha h, q} \bigg(\sideset{}{'} \sum_{j=0}^{n_{pl}} c_q^j \mathbf{t}_q^j \bigg) \Bigg) + |\mathrm{d\Omega}| \sum_{p=1}^{N_p} \sum_{q \in K_{\Omega_p}} \bigg(\frac{1}{4 \pi} \mathbf{\nabla}_{\alpha h} \phi \Big|_q  \nonumber \\
&\times& \mathbf{\nabla}_{\beta h} \phi \Big|_q  + \frac{1}{2} \delta_{\alpha \beta} \big(b_q - \rho_q \big) \phi_q \bigg) + |\mathrm{d\Omega}| \sum_{p=1}^{N_p} \sum_{I \in D^b_{p}} \sum_{q \in K_{\Omega_p}} \bigg( \mathbf{\nabla}_{\alpha h} b_I \Big|_q \big(\be_{\beta h,q} - \bR_{I,\beta} \big) \Big(\phi_q - \frac{1}{2} \, V_{I,q} \Big) - \frac{1}{2} \, \mathbf{\nabla}_{\alpha h} V_I \Big|_q \nonumber \\
&\times& \big(\be_{\beta h,q} - \bR_{I,\beta} \big) \, b_{I,q} - \frac{1}{2} \, b_{I,q} V_{I,q} \bigg) + \sigma_{\alpha \beta}^{E_c} \Bigg] \,,
\end{eqnarray}
where $|\mathrm{d\Omega}|$ denotes the volume associated with each grid point, $D_p^b$ is the set of all atoms in $\R^3$ whose pseudocharges have overlap with the processor domain $\Omega_p$, $\rho_q$ is the electron density at node $q$: 
\begin{eqnarray}
\rho_q = \frac{2}{|\mathrm{d\Omega}|} \sideset{}{'} \sum_{j=0}^{n_{pl}} c_q^j \rho_q^j \,, \quad \rho_q^j = \mathbf{w}_q^T \mathbf{t}_q^j \,,
\end{eqnarray}
${\bm \nabla}_h$ is the finite-difference approximation to the gradient with components $\nabla_{1h}$, $\nabla_{2h}$, and $\nabla_{3h}$, and $\be_{\beta h,q}$ is the spatial coordinate of the node $q$ in the $\beta$ direction. Note that in certain instances, only the pressure is needed and not the entire stress tensor.   In such cases, it is possible to compute the pressure by just taking the mean of the diagonal components of the stress tensor. However, a direct evaluation of the pressure---formulation presented in Appendix~\ref{App:Pressure}---is slightly more efficient and does not involve second order derivatives of the density matrix, which can be beneficial in certain other real-space discretization schemes such  as finite elements \cite{pask2005femeth,Phanish2010} and their mesh-free counterparts \cite{suryanarayana2011mesh}.


\section{Results and discussion} \label{Sec:Results}
We have implemented the proposed formulation for the stress tensor in the SQDFT code \cite{suryanarayana2017sqdft}. We now verify the accuracy and convergence of the  developed framework and demonstrate its practical utility by calculating the viscosity of hydrogen under extreme conditions from quantum molecular dynamics (QMD). In all simulations, we use the Gauss SQ method to evaluate the electron density during the self-consistent field (SCF) iteration \cite{Phanish2012,suryanarayana2013spectral},  and the Clenshaw-Curtis SQ method to evaluate the Hellmann-Feynman atomic forces \cite{pratapa2016spectral,suryanarayana2017sqdft}. In addition, we employ a twelfth-order accurate finite-difference discretization, the Periodic Pulay method \cite{banerjee2016periodic} for acceleration of the self-consistent field (SCF) iteration, and the AAR linear solver \cite{pratapa2016anderson,suryanarayana2019alternating} for calculation of the electrostatic potential as well as for application of the real-space Kerker preconditioner \cite{kumar2020preconditioning}.  Note that though we focus on high temperature calculations in the following, the proposed approach is equally applicable to calculations at ambient temperature, as shown in Appendix~\ref{App:LowT}. 

\subsection{Accuracy and convergence} \label{Subsec:AccCon}
We verify the accuracy and convergence of the proposed formulation and implementation  by considering three representative systems with atoms randomly perturbed by up to $15 \%$ of the equilibrium  interatomic distance: (i) a $32$-atom cell of face-centered cubic aluminum (Al) with lattice constant $7.78$ bohr, smearing $\sigma=4$ eV, and LDA exchange-correlation \cite{Kohn1965}; (ii) a $64$-atom cell of lithium hydride (LiH) with lattice constant $7.37$ bohr, smearing $\sigma=4$ eV, and LDA exchange-correlation; and (iii) a $64$-atom cell of diamond cubic carbon (C) with lattice constant $4.61$ bohr, smearing $\sigma=21.5$ eV, and GGA exchange-correlation \cite{perdew1996generalized}. For the aluminum and lithium hydride systems, we use Troullier-Martins pseudopotentials \cite{Troullier}, whereas for the carbon system, we use an ONCV pseudopotential \cite{hamann2013optimized}. 

\begin{figure}[h!]
\centering
\subfloat[Convergence with $n_{pl}$]{\includegraphics[keepaspectratio=true,width=0.4\textwidth]{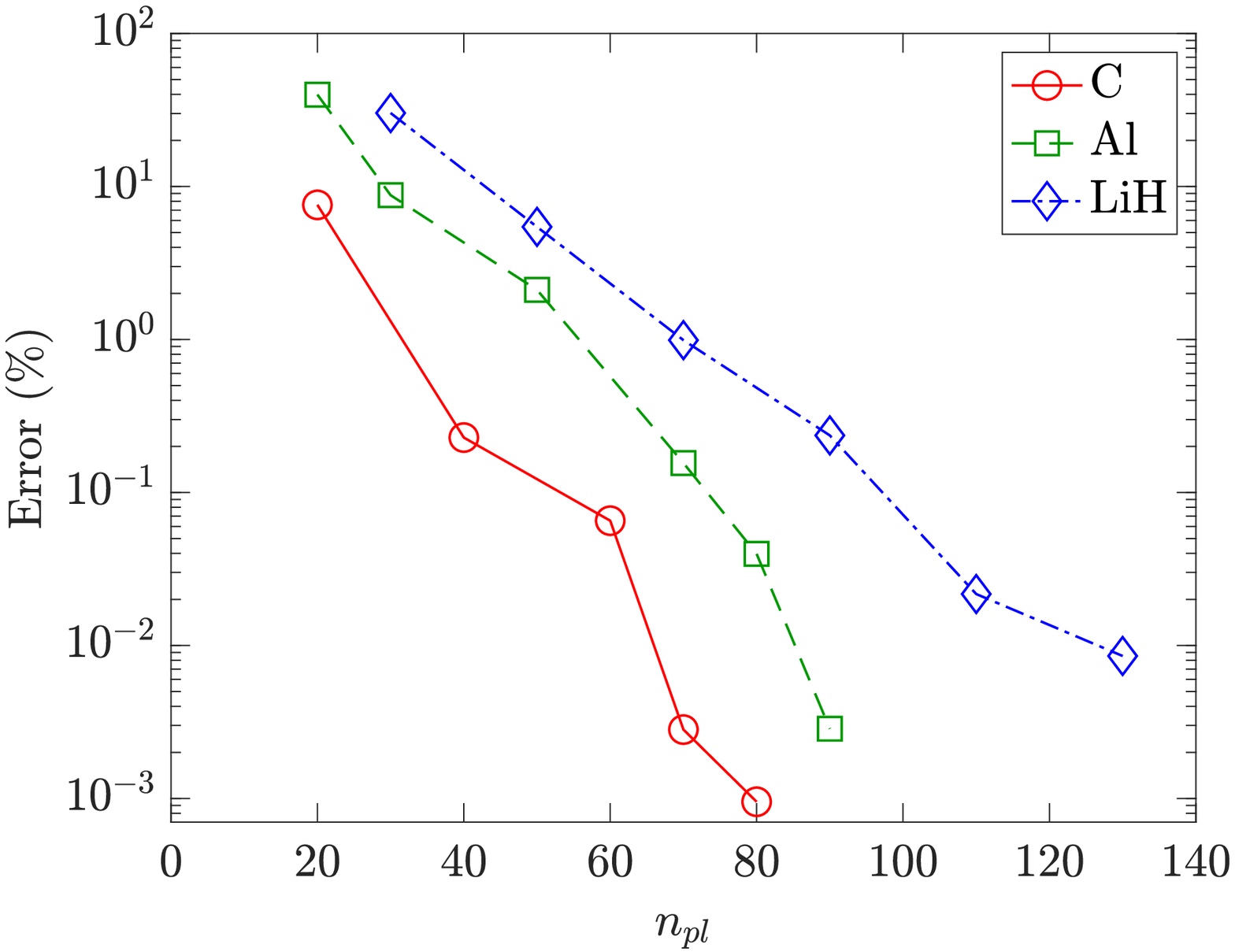} \label{Fig:errorVsnpl} } 
\subfloat[Convergence with $R_{cut}$]{\includegraphics[keepaspectratio=true,width=0.4\textwidth]{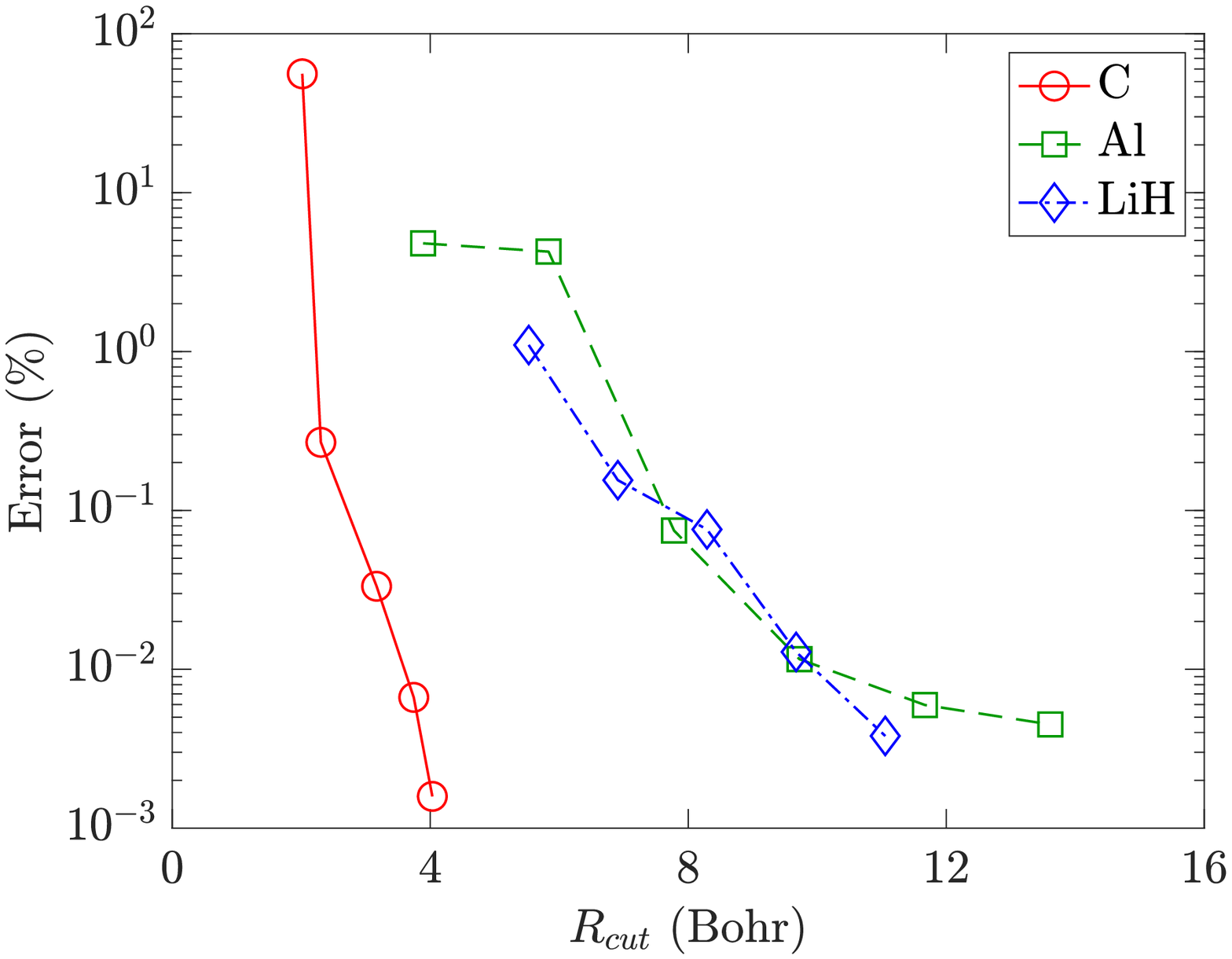} \label{Fig:errorVsrcut}}
\caption{\label{Fig:consistencyplots} Convergence of the stress tensor with respect to quadrature order $n_{pl}$ and truncation radius $R_{cut}$ for aluminum, lithium hydride, and carbon systems.  The error is defined to be the maximum difference in any component from the corresponding results obtained by SPARC.}
\end{figure}

First, we verify the convergence of  the stress tensor  with respect to the two additional parameters introduced by the SQ method, i.e., order of quadrature $n_{pl}$ and truncation radius $R_{cut}$.  We choose mesh-sizes of $h = 0.65$ bohr, $0.46$ bohr, and $0.23$ bohr for the aluminum, lithium hydride, and carbon systems, respectively. The values of $R_{cut}$ and $n_{pl}$ are chosen to be large enough so as to not influence the convergence behavior of $n_{pl}$ and $R_{cut}$, respectively. We present the results so obtained in Fig.~\ref{Fig:consistencyplots}, with the reference corresponding to diagonalization-based values obtained by SPARC \cite{xu2020sparc,ghosh2016sparc2} at the same mesh-size and with a $4 \times 4 \times 4$ Monkhorst-Pack grid for Brillouin zone integration.  The proposed formulation shows exponential convergence in the stress tensor  with respect to both $n_{pl}$ and $R_\text{cut}$. In particular, $\{n_{pl}, R_{cut}\} \sim \{55,6\}$, $\{n_{pl}, R_{cut}\} \sim \{70,5.6\}$, and $\{n_{pl}, R_{cut}\} \sim \{30,2.3\}$ are sufficient to obtain $1 \%$ accuracy in the stress tensor---errors typical in production simulations---for the aluminum, lithium hydride, and carbon systems, respectively. Note that the convergence with $R_{cut}$ is dictated by the smearing \cite{suryanarayana2017nearsightedness}, whereas convergence with $n_{pl}$ is dependent on the smearing, mesh-size, and the location of the Fermi level \cite{suryanarayana2013spectral}. 

\begin{figure}[h!]
\centering
\includegraphics[keepaspectratio=true,,width=0.4\textwidth]{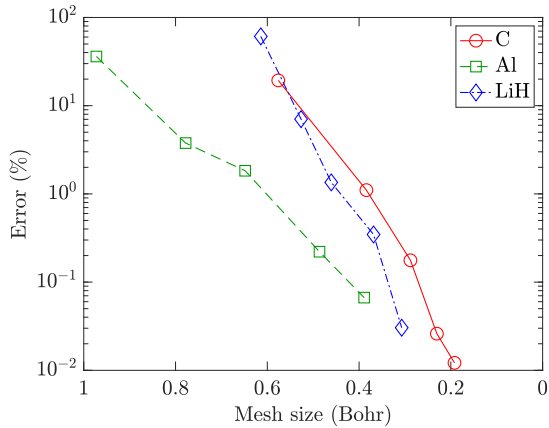}
\caption{\label{Fig:convergenceplot} Convergence of the stress tensor with mesh size for aluminum, lithium hydride, and carbon systems. The error is defined to be the maximum difference  in any component 
from the corresponding results obtained by ABINIT.}
\end{figure}

Next, we verify convergence of the stress tensor with respect to spatial discretization, using  highly converged diagonalization-based results from the planewave code ABINIT  \cite{ABINIT} as reference. Specifically, we utilize $\{n_{pl},R_{cut}\} = \{150,12\} $, $\{200,12\}$, and $\{180,6\}$ for the aluminum, lithium hydride, and carbon systems, respectively. These parameters are sufficient to put the associated errors well below the mesh errors of interest, as evident from the results in  Fig.~\ref{Fig:consistencyplots}. In ABINIT, we employ a planewave cutoff of $60$ Ha and a $4 \times 4 \times 4$ Monkhorst-Pack grid for Brillouin zone integration, which translates to stresses that are converged to within $0.01 \%$. We present the results so obtained in Fig.~\ref{Fig:convergenceplot}, from which it is clear that there is systematic convergence in the stress tensor computed using the proposed approach. In particular, stresses accurate to within $0.1 \%$ are readily obtained.

\subsection{High temperature QMD: viscosity of hydrogen at a million kelvin}

Hydrogen is the most abundant element in the universe and its properties have important implications in astrophysics and planetary science, including inertial confinement fusion (ICF) experiments where isotopes deuterium and tritium serve as the fuel.  In particular, material properties such as the mass diffusivity, thermal conductivity, and viscosity of warm dense hydrogen impact the onset of convection or turbulence in a hydrodynamic system which in turn impacts the distribution of energy in a star, planet, or ICF capsule. These important quantities are very difficult to measure under extreme conditions of pressure and temperature. However, they can be accurately calculated from first principles QMD simulations. 

In this work, we calculate the viscosity of hydrogen at a temperature of 10$^6$ K and density of  2 g/cm$^3$. To do so, we consider a canonical ensemble of 64 hydrogen atoms and employ the LDA exchange-correlation functional, a local ONCV pseudopotential suitable for the target temperature, and the isokinetic Gaussian thermostat \cite{NVK}. In order to extract a statistical uncertainty of the final result, we average over QMD simulations corresponding to 10 different initial conditions for the atom positions and velocities,  where each simulation has been run for more than 25000 steps with a  time step of 0.04 fs. Since macroscopic dynamics properties can be written as the time-integral of a microscopic time correlation function using Green-Kubo (GK) relations~\cite{hansen2013theory}, we calculate the viscosity using the relation: 
\begin{equation}
\eta (t) = \frac{|\Omega|}{k_B T} \int_0^t  \left( \frac{1}{5} \sum_{i=1}^{5}  \langle s_i(t') s_i(0)\rangle \right) \, \mathrm{dt'} \,,
\label{Eqn:viscosity}
\end{equation} 
where $k_B$ is the Boltzmann constant, $T$ is the temperature, $\langle \cdot \rangle$ denotes the ensemble average, and $s_i$ are the independent components of the deviatoric (i.e., traceless) stress tensor, i.e., $\sigma_{12}$, $\sigma_{23}$, $\sigma_{31}$,  $(\sigma_{11}-\sigma_{22})/2$, and $(\sigma_{22}-\sigma_{33})/2$ \cite{Alfe1998}. 

\begin{figure}
\includegraphics[keepaspectratio=true,,width=0.7\textwidth]{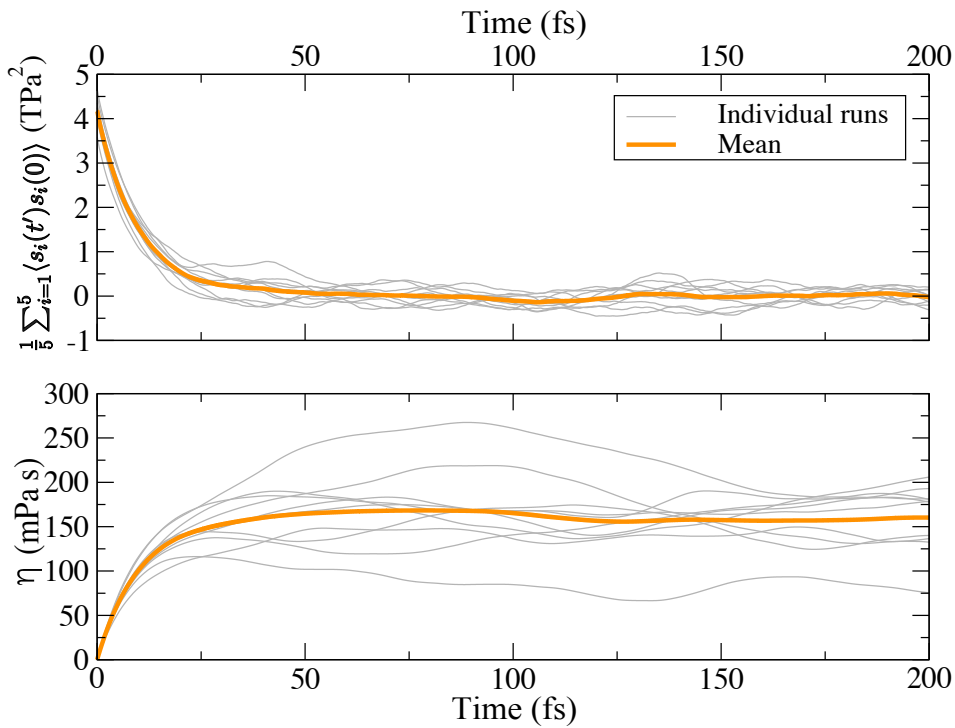}
\caption{\label{Fig:viscosityplot} Ensemble average (top) and viscosity (bottom) for  hydrogen at $10^6$~K and density 2 g/cm$^3$.}
\end{figure}

\begin{figure}
\includegraphics[keepaspectratio=true,,width=0.4\textwidth]{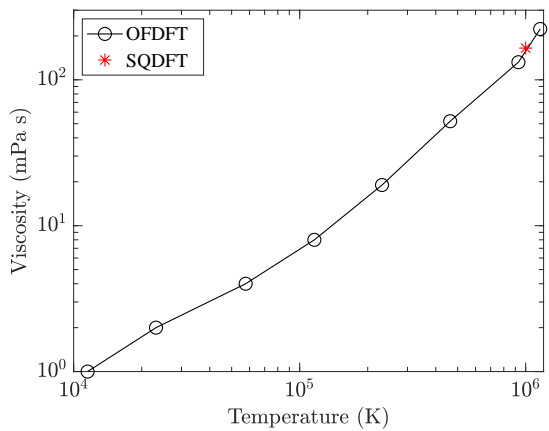}
\caption{\label{Fig:visco_vs_ofdft} Comparison between OFDFT \cite{Sjostrom2015} and SQDFT for the viscosity of hydrogen at a density of 2 g/cm$^3$.}
\end{figure}

We present the results so obtained in Fig.~\ref{Fig:viscosityplot}. It is clear that the mean of the ensemble average for the independent components of the deviatoric stress tensor  decays in around 50 fs, resulting in  a viscosity of 164 $\pm$ 39 mPa s. This full Kohn-Sham DFT result is consistent with recent OFDFT calculations by Sjostrom and Daligault~\cite{Sjostrom2015} (Fig~\ref{Fig:visco_vs_ofdft}), while being free of the kinetic energy functional approximation inherent to OFDFT.

\section{Concluding remarks} \label{Sec:Conclusions}
In this work, we have presented an accurate and efficient real-space formulation for computation of the Hellmann-Feynman stress tensor in $\mathcal{O}(N)$  Kohn-Sham DFT calculations. While applicable at any temperature, the formulation is most efficient at high temperature where the Fermi-Dirac distribution becomes smoother and density matrix becomes correspondingly more localized. After rewriting the orbital-dependent real-space stress tensor in terms of the density matrix, we developed an approach for evaluating it using the infinite-cell Clenshaw-Curtis variant of the Spectral Quadrature (SQ) method. Notably, the developed framework is applicable to metallic as well as insulating systems, is highly parallelizable, becomes more efficient as the temperature is increased, and   can be used to compute stresses corresponding to the infinite crystal without need of Brillouin zone integration. We demonstrated that the resulting formulation  converges systematically with respect to both polynomial order and localization radius to the exact diagonalization result, and with respect to mesh size to established planewave results. Finally, we employed the new formulation to compute the viscosity of hydrogen at a million kelvin from Kohn-Sham quantum molecular dynamics, where we found agreement with previous more approximate orbital-free density functional methods.  


\section*{Acknowledgements}
 This work was performed in part under the auspices of the U.S. Department of Energy by Lawrence Livermore National Laboratory under Contract DE-AC52-07NA27344. Support from the Advanced Simulation \& Computing / Physics \& Engineering Models program at LLNL is gratefully acknowledged. PS also acknowledges the support of the U.S. National Science Foundation (NSF) under Grant No. 1663244. Time on the Quartz supercomputer was provided by the Computing Grand Challenge program at LLNL. We thank Donald Hamann for use of and assistance with a development version of the ONCVPSP code.
 
\section*{Data Availability}
The data that support the findings of this study are available from the corresponding author upon reasonable request.


\appendix
\section{Pressure formulation, implementation, and verification}\label{App:Pressure}
In certain applications, such as the calculation of equations of state \cite{PhysRevB.99.165103}, only the pressure is required and not the entire stress tensor. Though the pressure can be computed from diagonal elements of the stress tensor, here we present a suitable reformulation that can not only improve the efficiency of the calculation, albeit slightly, but also make it more amenable to other real-space discretizations such as finite elements \cite{pask2005femeth,Phanish2010} and their mesh-free counterparts \cite{suryanarayana2011mesh}. We now present the reformulation for the pressure, its evaluation within the infinite-cell Clenshaw-Curtis Spectral Quadrature (SQ) method \cite{pratapa2016spectral} as implemented in the SQDFT code \cite{suryanarayana2017sqdft}, and its verification through selected examples. 

The expression for the pressure in terms of the density matrix can be written as: 
\begin{eqnarray}
P &=& -\frac{1}{3} \left( \sigma \mkern-2mu _{11}  + \sigma \mkern-2mu _{22} + \sigma \mkern-2mu _{33} \right) \nonumber \\
&=& -\frac{1}{3 |\Omega |} \Bigg[ -4  \int_{\Omega} \big(\mathcal{D} \mathcal{H}\big) (\be,\be) \, \mathrm{d \be}  + 3E_{xc}(\rho_\mathcal{D}, \bm{\nabla} \rho_\mathcal{D}) - \int_{\Omega} \Bigg( V_{xc} \big(\rho_\mathcal{D}(\be), \gradx \rho_\mathcal{D}(\be) \big) + \frac{\partial \varepsilon_{xc}\big(\rho_\mathcal{D}(\be),\gradx \rho_\mathcal{D}(\be) \big)}{\partial \big(\gradx \rho_\mathcal{D}(\be)\big)} \cdot \gradx \rho_\mathcal{D}(\be) \Bigg) \nonumber \\ 
& \times & \rho_\mathcal{D}(\be) \, \mathrm{d \be} - E_{nl}(\denM,\bR) - \, 4  \sum_J \sum_{lm} \gammaJl \,  \sum_{J'} \int_\Omega \int_\Omega \pchis(\be,\bRJp) \big(\be - \bRJp \big) \cdot \gradx \mathcal{D}(\be,\mathbf{y}) \tchins(\mathbf{y},\bRJ) \, \mathrm{d \be} \, \mathrm{d \mathbf{y}} \nonumber \\
&+& \frac{1}{4 \pi} \int_\Omega \big| \gradx \phi(\be,\bR) \big|^2 \, \mathrm{d \be} + \sum_I \int_\Omega \gradx b_I(\be,\bR_I) \cdot \big(\be - \bR_I\big) \Big(\phi(\be,\bR) - \frac{1}{2} \, V_I(\be,\bR_I) \Big) \, \mathrm{d \be} \nonumber \\
&-& \frac{1}{2} \sum_I \int_\Omega \gradx V_I(\be,\bR_I) \cdot \big(\be - \bR_I\big)\, b_I(\be,\bR_I) \, \mathrm{d \be} + \frac{1}{2} \int_\Omega \big(\rho_\mathcal{D}(\be) + 3 \, b(\be,\bR) \big) \phi(\be,\bR) \, \mathrm{d \be} - 3 E_{self}(\bR) + \sum_{i=1}^3 \sigma \mkern-2mu ^{E_c}_{ii} \Bigg]\,,  \label{Eqn:ON:Pressure}
\end{eqnarray} 
where the Hamiltonian
\begin{eqnarray}
\mathcal{H} = - \frac{1}{2} \nabla^2 + V_{xc} + \phi + V_{nl} 
\end{eqnarray} 
is introduced to eliminate the  kinetic energy contribution that involves explicit second order derivatives of the density matrix.

In the context of the infinite-cell Clenshaw-Curtis method described in Section~\ref{Sec:StressSQ}, the expression for the pressure takes the form: 
\begin{eqnarray}
P &=& -\frac{1}{3 |\Omega|} \Bigg[|\mathrm{d\Omega}| \sum_{p=1}^{N_p} \sum_{q \in K_{\Omega_p}} \Bigg( -\frac{4}{|\mathrm{d\Omega}|} \sideset{}{'} \sum_{j=0}^{n_{pl}} \big(\chi_q c_q^j + \zeta_q d_q^j \big) \rho_q^j + 3 \, \epsilon_{xc}(\rho_q) \rho_q - V_{xc}(\rho_q) \rho_q - \rho_q \frac{\partial \varepsilon_{xc}\big(\rho,\bm \nabla_{h} \rho \big)}{\partial \big(\bm \nabla_{h} \rho \big)} \Bigg|_q \cdot \bm \nabla_{h} \rho \Big|_q \Bigg) \nonumber \\
&-& 2 \sum_{p=1}^{N_p} \sum_{I \in D^c_{p'}} \sum_{q \in K_{\Omega_p}} \mathbf{w}_q^T V_{nl,q}^I\Bigg(\sideset{}{'} \sum_{j=0}^{n_{pl}} c_q^j \mathbf{t}_q^j + 2 \, \big(\mathbf{X}_{h,q} - \bR_I \mathbf{I} \big) \cdot \nabla _{h, q} \bigg(\sideset{}{'} \sum_{j=0}^{n_{pl}} c_q^j \mathbf{t}_q^j \bigg) \Bigg) + |\mathrm{d\Omega}| \sum_{p=1}^{N_p} \sum_{q \in K_{\Omega_p}} \bigg(\frac{1}{4 \pi} \Big| {\bm \nabla}_{h} \phi \big|_q \Big|^2 \nonumber \\
&+ & \frac{1}{2} \big(\rho_q + 3 \, b_q \big) \phi_q \bigg) + |\mathrm{d\Omega}| \sum_{p=1}^{N_p} \sum_{I \in D^b_{p}} \sum_{q \in K_{\Omega_p}} \bigg( {\bm \nabla}_{h} b_I \Big|_q \cdot \big(\be_{h,q} - \bR_I\big) \Big(\phi_q - \frac{1}{2} \, V_{I,q} \Big) - \frac{1}{2} {\bm \nabla}_{h} V_I \Big|_q .\big(\be_{h,q} - \bR_I \big) \, b_{I,q} - \frac{3}{2} \, b_{I,q} V_{I,q} \bigg) \nonumber \\
&+& \sum_{i=1}^3 \sigma_{ii}^{E_c} \Bigg] \,,
\end{eqnarray}
where the derivation for the first term can be found in previous work \cite{pratapa2016spectral}. In particular, rather than performing a Chebyshev polynomial expansion for the density matrix, it is done for the product of the Hamiltonian with the density matrix. This  bypasses the need to calculate second-order derivatives of the density matrix, thereby  making the evaluation of the pressure slightly more efficient as well as more amenable to other real-space discretization schemes.

\begin{figure}[h!]
\centering
\subfloat[Convergence with $n_{pl}$]{\includegraphics[keepaspectratio=true,width=0.4\textwidth]{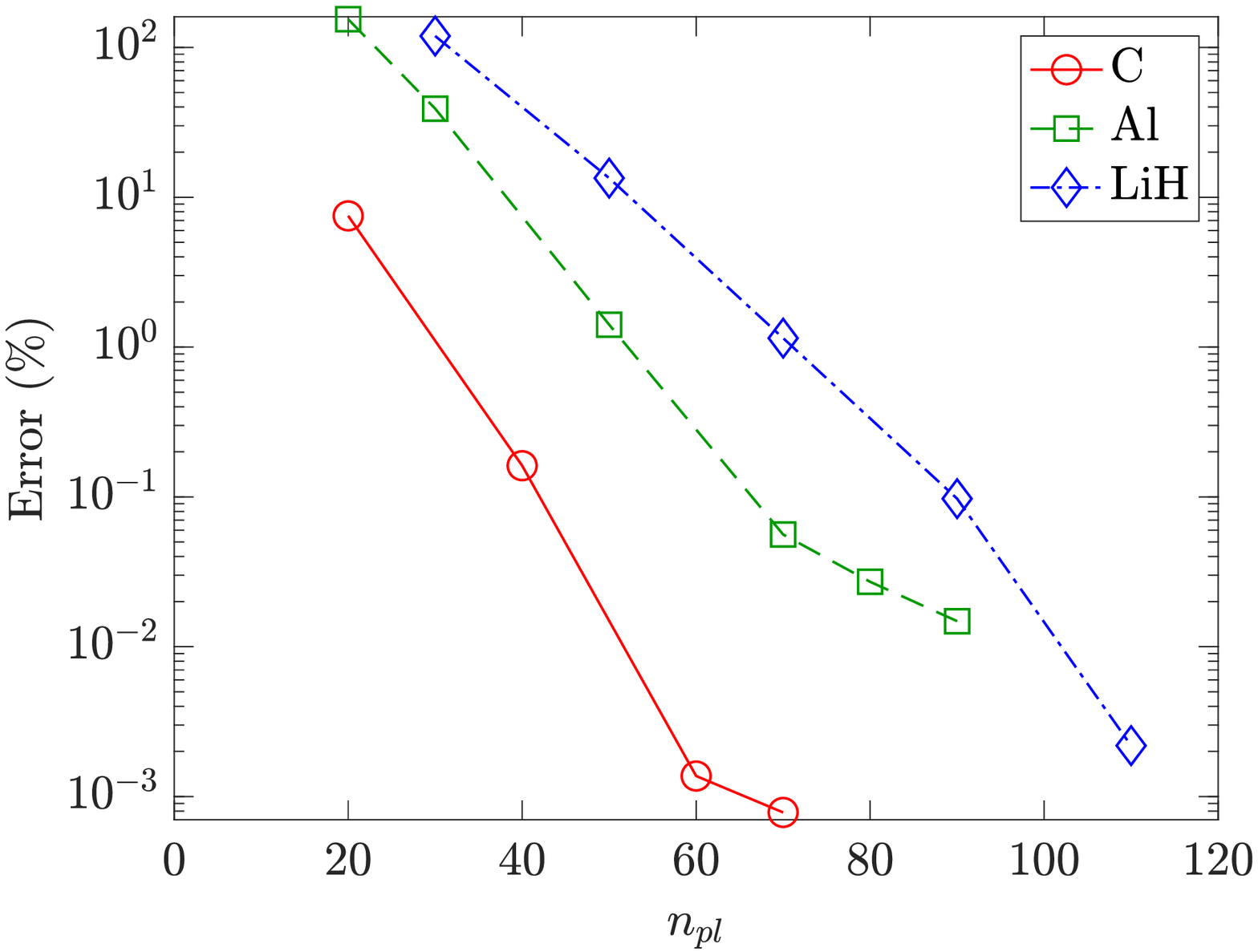} \label{Fig:errorVsnpl_p} } 
\subfloat[Convergence with $R_{cut}$]{\includegraphics[keepaspectratio=true,width=0.4\textwidth]{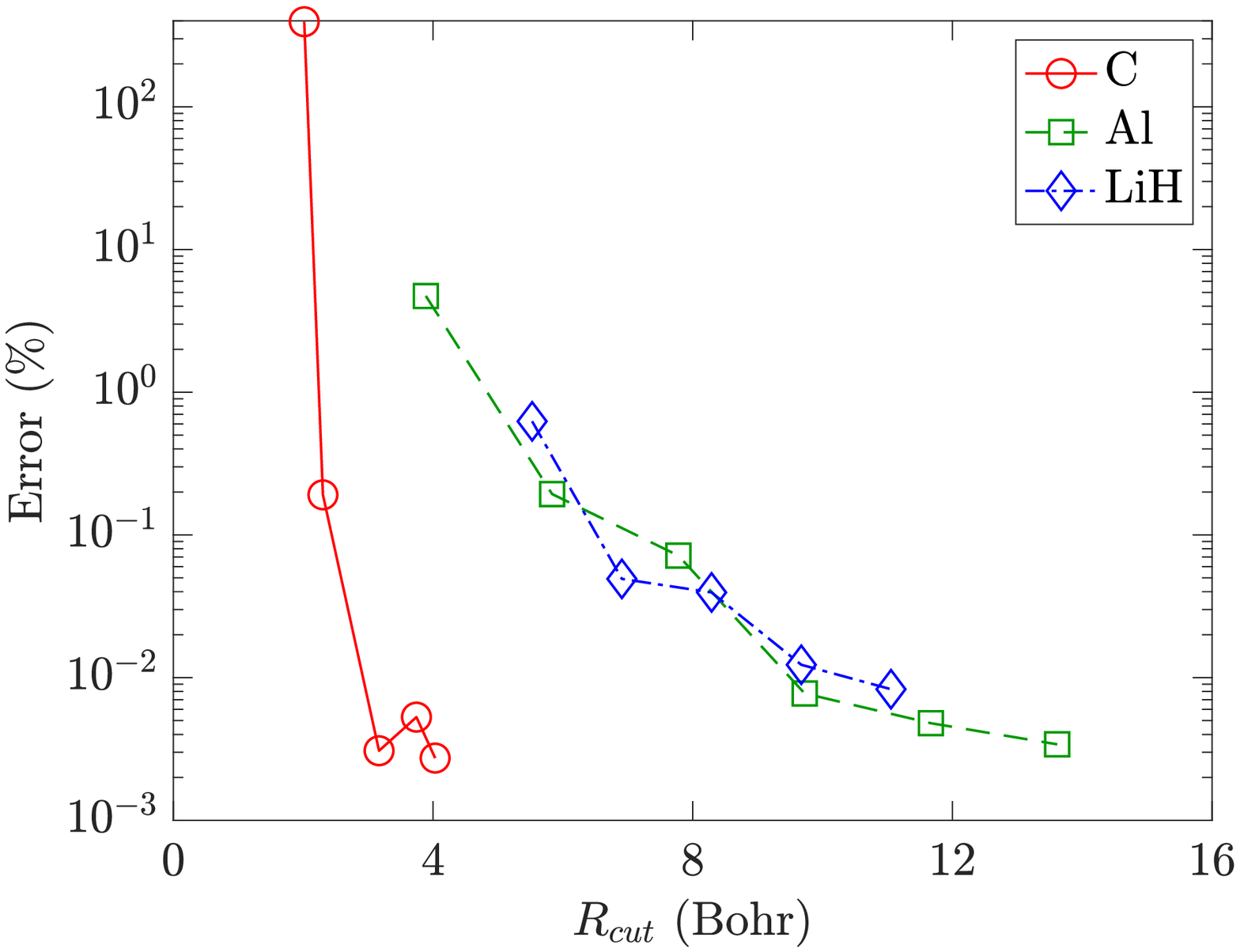} \label{Fig:errorVsrcut_p}} \\ 
\subfloat[Convergence with mesh size]
{\includegraphics[keepaspectratio=true,width=0.4\textwidth]{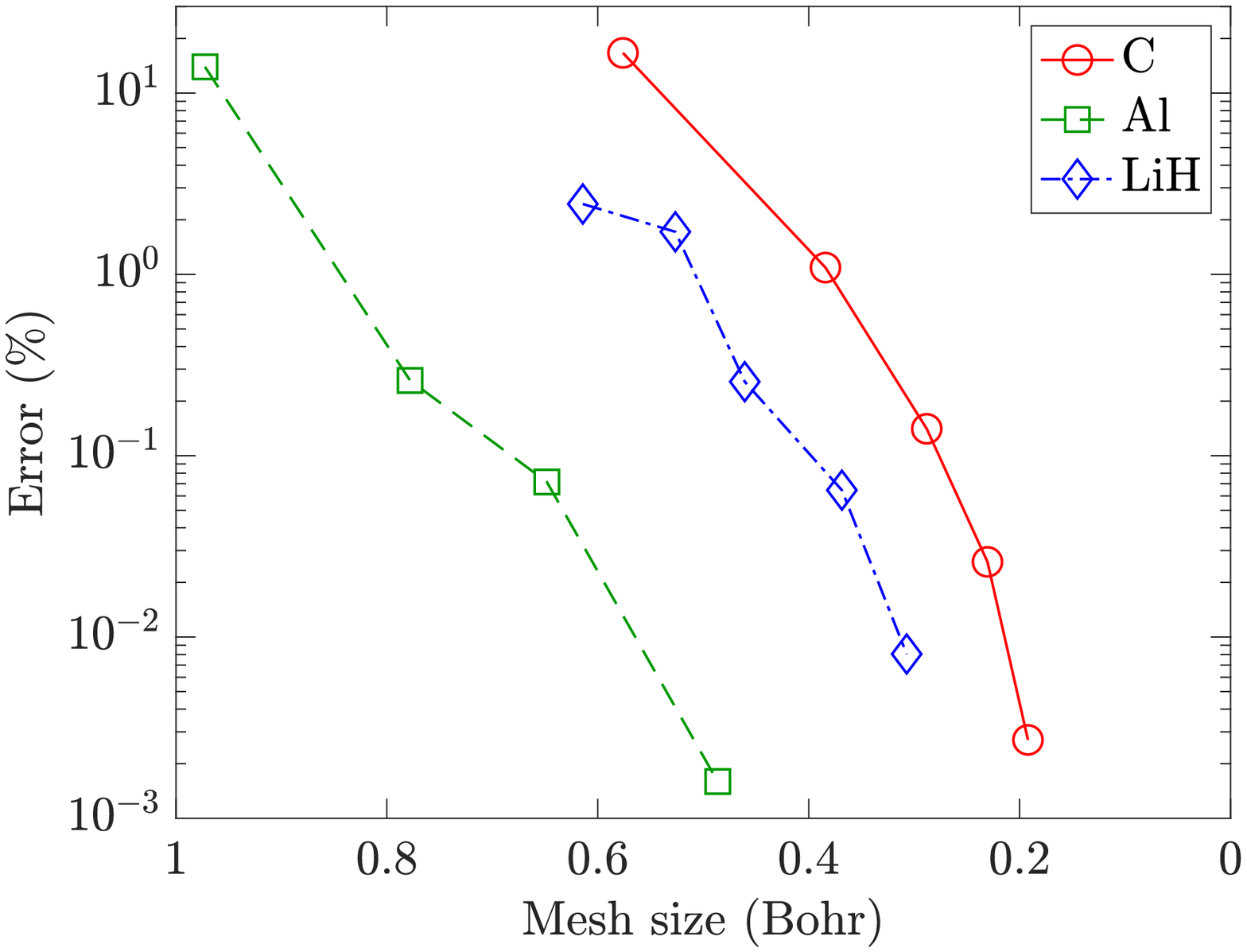} \label{Fig:errorVsrcut_mesh_p}}
\caption{\label{Fig:AccConv:P} Convergence of the pressure with respect to quadrature order $n_{pl}$, truncation radius $R_{cut}$, and mesh size for aluminum, lithium hydride, and carbon systems.  The error in (a) and (b) is defined to be the difference from the corresponding results obtained by SPARC, whereas in (c) it is the difference from the corresponding results obtained by ABINIT.}
\end{figure}

We now verify the accuracy and convergence of the proposed formulation for the pressure using the same examples as in Section~\ref{Subsec:AccCon}, where a detailed description of the tests can also be found. First, we check convergence of  the pressure with respect to  quadrature order $n_{pl}$ and truncation radius $R_\text{cut}$ in Figs.~\ref{Fig:errorVsnpl_p} and \ref{Fig:errorVsrcut_p}, respectively, with the values from the real-space code SPARC \cite{xu2020sparc,ghosh2016sparc1} again used as reference.   It is clear that there is exponential convergence in the pressure  with respect to both parameters, the curves being very similar to those obtained for the stress tensor. Next, we check convergence with spatial discretization in Fig.~\ref{Fig:errorVsrcut_mesh_p}, 
the results from the planewave code ABINIT \cite{ABINIT} again used as reference. We observe that there is systematic convergence in the pressure, similar to the stress tensor. Note that the convergence is faster for the pressure, likely a consequence of the stress tensor's off-diagonal components containing mixed derivatives, which are found to generally cause slower convergence with discretization \cite{AbhirajKP}.

\section{Verification of stress tensor formulation at ambient conditions} \label{App:LowT}
\begin{figure}[h!]
\centering
\subfloat[Convergence with $n_{pl}$]{\includegraphics[keepaspectratio=true,width=0.4\textwidth]{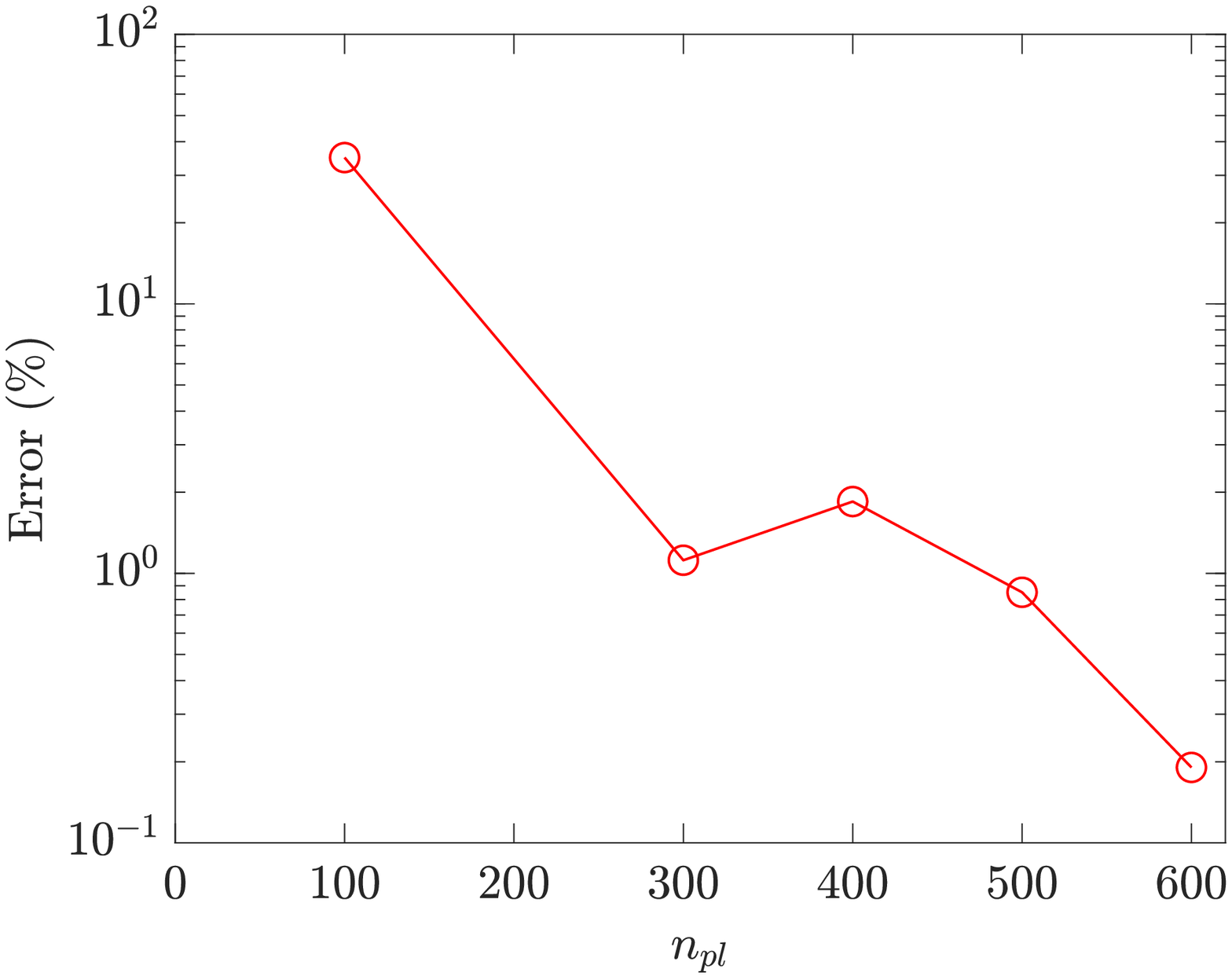} \label{Fig:errorVsnpl_lowT} } 
\subfloat[Convergence with $R_{cut}$]{\includegraphics[keepaspectratio=true,width=0.4\textwidth]{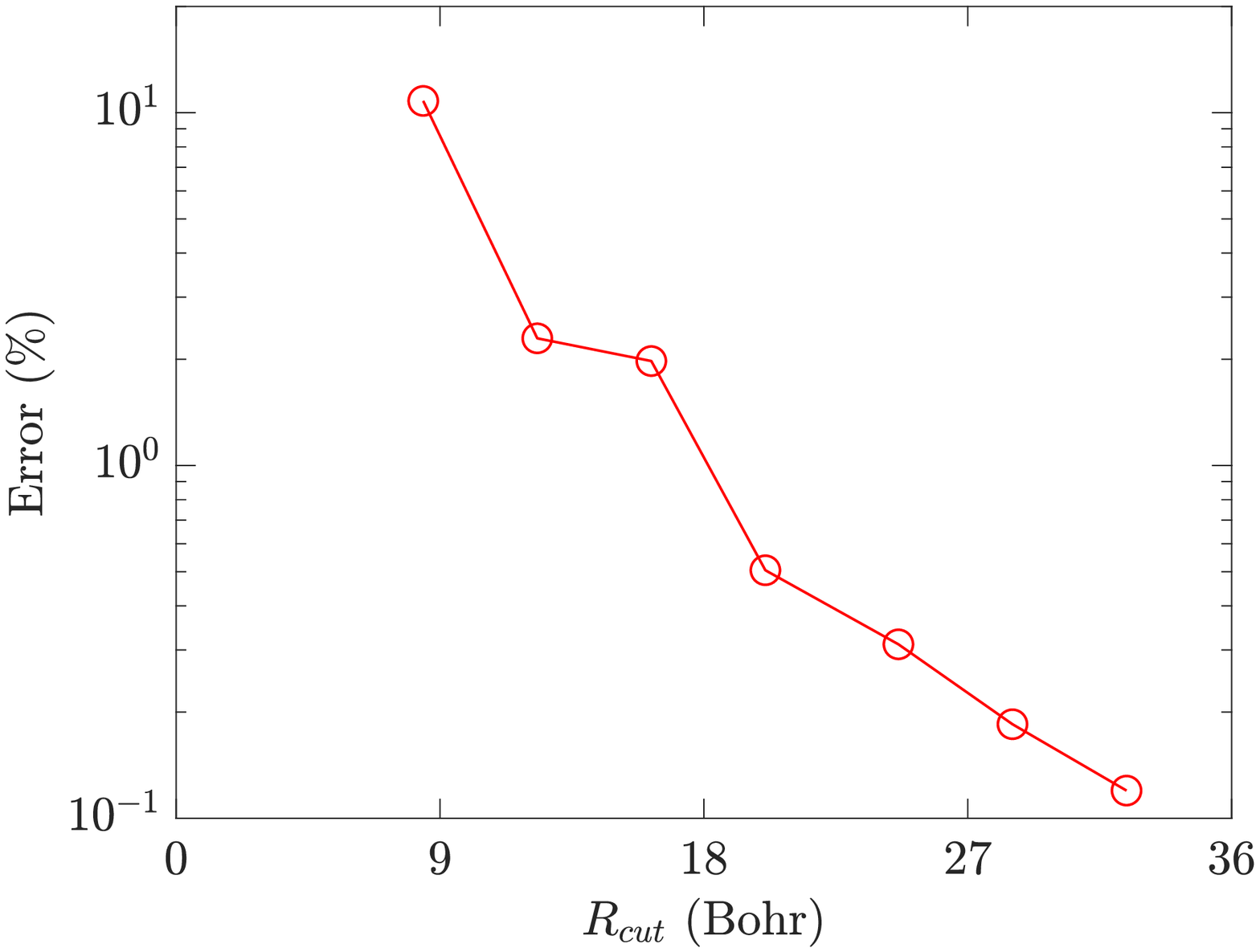} \label{Fig:errorVsrcut_lowT}}
\caption{\label{Fig:consistencyplots_lowT} Convergence of the stress tensor with respect to quadrature order $n_{pl}$ and truncation radius $R_{cut}$ for aluminum at ambient conditions.  The error in the stress tensor is defined to be the maximum difference  in any component from the corresponding results obtained by SPARC.}
\end{figure}

We now demonstrate the accuracy of the proposed stress tensor formulation and implementation for Kohn-Sham calculations at ambient temperature. Specifically, we consider a randomly perturbed $4$-atom  aluminum system and choose a smearing of  $\sigma = 0.27$ eV, typical of the values employed for metallic systems at ambient conditions to facilitate self-consistent convergence \cite{VASP,ABINIT}.  In Fig.~\ref{Fig:consistencyplots_lowT}, we verify the convergence of the stress tensor with respect to $n_{pl}$ and $R_\text{cut}$ for mesh size $h=0.65$ bohr, with the reference diagonalization results obtained by SPARC at the same mesh-size and a $8 \times 8 \times 8$ Monkhorst-Pack grid for Brillouin zone integration.  We employ $R_{cut}=28$ bohr and $n_{pl}=650$ for the convergence with $n_{pl}$ and $R_{cut}$, respectively. As observed previously for larger smearing values, there is exponential convergence with both parameters. Note that larger values of $n_{pl}$ and $R_{cut}$ are required due to the reduced smoothness of the Fermi-Dirac function and the decreased locality of the electronic interactions, respectively. 

It is worth noting that for the system chosen here, Brillouin zone integration is critical for the accurate calculation of the stress tensor. For example, there is a $29$\% error in values of the stress when a $\Gamma$-point calculation is performed using ABINIT/SPARC. This verifies the accuracy of the infinite-cell SQ method in calculating the stress tensor corresponding to the infinite crystal, without the need for Brillouin zone integration or large supercells. Also note that we do not show convergence with mesh size here, since given large enough $n_{pl}$ and $R_{cut}$, the results are independent of temperature/smearing, and therefore the same curves as in Figs.~\ref{Fig:convergenceplot}  are obtained. Overall, it can be concluded that the proposed stress tensor formulation and implementation are not restricted by the choice of smearing/temperature. 

\end{document}